\documentclass[cpp,a4paper,fleqn]{w-art}

\usepackage{times,cite,w-thm}
\theoremstyle{plain}

\theoremstyle{definition}

\usepackage[]{graphicx}

\usepackage{subfigure}
\usepackage[english]{babel}
\usepackage{sidecap}

\newcommand{\ket}[1]{{\ensuremath  \bigl | \,  #1  \, \bigr \rangle }}
\newcommand{\bra}[1]{{\ensuremath \bigl \langle \,  #1 \, \bigr | }}

\begin{document}

\DOIsuffix{theDOIsuffix}

\Volume{XX}
\Month{XX}
\Year{XXXX}

\pagespan{1}{}

\Receiveddate{XXXX}
\Reviseddate{XXXX}
\Accepteddate{XXXX}
\Dateposted{XXXX}

\keywords{photoionization, quantum kinetic theory, nonequilibrium Greens functions, multiconfiguration time-dependent Hartree-Fock}
\subjclass[pacs]{05.30.-d, 32.80.Fb}

\title[Photoionization of atoms]{Quantum kinetic approach to time-resolved \\photoionization of atoms}
\author[M. Bonitz]{M. Bonitz
  \footnote{Corresponding author\quad E-mail:~\textsf{bonitz@physik.uni-kiel.de},
            Phone: +\,49\,431\,880-4122,
            Fax: +\,49\,431\,880-4094}}
\author[D. Hochstuhl]{D. Hochstuhl}
\author[S. Bauch]{S. Bauch}
\author[K. Balzer]{K. Balzer}
\address{Institute for Theoretical Physics and Astrophysics, Christian Albrechts University Kiel, \\ Leibnizstrasse 15, D-24098 Kiel, Germany}

\begin{abstract}
Theoretical approaches to the photoionization of few-electron atoms are discussed. These include nonequilibrium Greens functions 
and wave function based approaches. In particular, the Multiconfiguration Time-Dependent Hartree-Fock method is discussed and applied 
to a model one-dimensional atom with four electrons. We compute ground state energies and the time-dependent photoionization by the field a strong laser 
pulse with two different frequencies in the ultraviolet. 
\end{abstract}

\maketitle

\section{Introduction}
In recent years ultrashort, high-harmonic generated vacuum and extreme ultraviolet (vuv/xuv) laser pulses have become available. They open the way
towards time-resolved observation of electronic dynamics in plasmas, atoms and condensed matter, for an overview see e.g. \cite{Bauer2005}. 
Electronic motion in atoms and relaxation processes can now be studied experimentally on sub-femtosecond time scales, e.g. \cite{Corkum2007,Drescher2001,Uiberacker2007}, for a recent overview see \cite{krausz2009}. This creates the need for complementary theoretical investigations.
Interaction of laser radiation with matter has been actively studied in recent years using a quantum kinetic theory which was derived using nonequilibrium Greens functions (NEGF) methods, e.g. \cite{kremp-etal.99pre,bonitz-etal.99cpp,haberland-etal.01pre}. In contrast, application of NEGF to single atoms has only recently been attempted \cite{dahlen07,hochstuhl_physe09}. While the ground state properties were very well reproduced, time-dependent calculations describing the ionization dynamics of atoms during and after a laser pulse turned out to be computationally very expensive \cite{hochstuhl_physe09}. 

In this paper we give a brief overview on the application of NEGF to time-resolved laser-atom interaction and outline the current problems. Furthermore, we discuss an alternative approach which is based on the $N-$electron wave function -- the Multiconfiguration Time-Dependent Hartree-Fock method (MCTDHF). The latter is quite efficient in application to few-electron atoms and small molecules. This is demonstrated by numerical results for a four-electron one-dimensional model atom subject to a short laser pulse.

\section{Theoretical concepts}\label{s:theory}
The unperturbed hamiltonian of $N$ electrons in the atom is given by (we use atomic units) 
\begin{equation}\label{h0}
\hat H_0  \ = \ \sum_{i=1}^{N} \left\{-\frac{\nabla_i^2}{2} + v(\mathbf r_i) \right\} + \frac 12 \sum_{i\neq j}^{N} w(\mathbf r_i - \mathbf r_j)\,,
\end{equation}
where the potential of the nucleus, $v(\mathbf r_i)$, as well as the two-particle Coulomb interaction $w(\mathbf r_i - \mathbf r_j)= |\mathbf r_i - \mathbf r_j|^{-1}$ are assumed to be spin-independent. 
For times $t>0$, the atom is disturbed by a time-dependent external field, and the hamiltonian is modified:
\begin{equation}
\hat H(t) \ = \ \hat H_0 \, - \, {\bf E}(t)\sum_{i=1}^{N}\mathbf r_i\,,
\label{h}
\end{equation}
where the electromagnetic field is described in dipole approximation, and the field envelope is assumed to have a Gaussian shape 
with a pulse duration $\tau$ and $t_0\gg \tau$
\begin{align}\label{eq:gauss}
E(t) \ = \ {\cal E}_0 \cos \bigl[\omega (t-t_{\text{0}}) \bigr] \exp \left[- \frac{(t-t_{\text{0}})^2}{2\tau^2}\right] \,.
\end{align}
%

\subsection{Quantum kinetic theory for a statistical ensemble of atoms}
To treat the interaction of an ensemble of atoms with the electromagnetic field a statistical theory is required. It can be based on density operators \cite{bonitz-book} or nonequilibrium Greens functions. Here we discuss the latter concept.
Starting point are the fermionic field operators of the electrons (the nuclei are assumed to be fixed)
$\psi^{\dagger}$ und $\psi$, which guarantee that the spin statistics theorem is obeyed,
\begin{eqnarray*}
\psi(1)\psi(2) +
\psi(2)\psi(1)
&=&
\psi^{\dagger}(1)\psi^{\dagger}(2)
+ \psi^{\dagger}(2)\psi^{\dagger}(1)=0,
\nonumber\\
\psi(1)\psi^{\dagger}(2)
+ \psi^{\dagger}(2)\psi(1)
&=& \delta(1-2),
\label{psi-comm}
\end{eqnarray*}
where $t_1=t_2$ and we denoted $1\equiv ({\bf r}_1,t_1,s_{z1})$ [spin indices will be suppressed in the following]. 
The nonequilibrium properties of the electrons are completely determined by the two-time correlation functions 
(averaging is over the $N$-particle 
density operator of the system)
\begin{eqnarray}
g^>(1,1')=\frac{1}{i}\langle\psi(1)\psi^{\dagger}(1')\rangle\,,
\qquad g^<(1,1')=
-\frac{1}{i}\langle\psi^{\dagger}(1')\psi(1)\rangle,
\label{ggtls_def}
\end{eqnarray}
The two independent functions $g^{>}$ and $g^{<}$ contain the complete statistical information, in particular the nonequilibrium one-body density matrix, according to 
\begin{equation}
\rho({\bf r}_1,{\bf r}_1',t) = -i g^{<}(1,1')|_{t_1=t'_1}.
\label{rho-def}
\end{equation}
The equations of motion of $g^{>}$ and $g^{<}$ are the Keldysh/Kadanoff-Baym equations (KBE), e.g. \cite{kb-book,kremp-book} which are the most general starting point for developing a quantum statistical theory of matter in a strong electromagnetic field.
In recent years the KBE have been successfully solved numerically for spatially homogeneous plasmas, e.g. \cite{bonitz-etal.jpcm96,bonitz97} and 
for a weakly inhomogeneous electron plasma \cite{kwong-etal.00prl}. 

Yet these solutions require to propagate the correlation functions in the two-time plain which is computationally very expensive. Therefore, approximate methods have been developed. One possibility consists in expressing the correlation functions by the simpler single-particle distribution function $f(t)$ [or density matrix $\rho(t)$, Eq.~(\ref{rho-def})] via the 
so-called generalized Kadanoff-Baym ansatz \cite{lipavski}. For plasmas interacting with a strong field this ansatz has been generalized in refs.~\cite{kremp-etal.99pre,bonitz-etal.99cpp}. The emerging quantum kinetic equation for the Wigner distribution which includes nonlinear field effects, 
inverse bremsstrahlung, collisional absorption and harmonic generation has been solved in ref. \cite{haberland-etal.01pre}. The second direction of 
recent research in the kinetic theory of laser-plasma interation was devoted to include neutral atoms and their excitation and ionization. Starting from the 
KBE atomic Bloch equations were derived \cite{semkat-etal.05jp,kremp-etal.06jp} which allow to compute the absorption spectrum of an ensemble of atoms in thermodynamic equilibrium. This approach is very useful for weak excitation near the linear response regime. 

However, in the case of strong field ionization of atoms we expect essential nonequilibrium dynamics effects related to ionization of several electrons and  electron-electron correlations developing rapdily on femtosecond and even sub-femtosecond time scales. These phenomena which have already been observed in experiments require a fully nonequilibrium approach. Starting again from the KBE we have in recent years developed a quantum kinetic approach for the electron dynamics in a single atom. The main difference to the previous applications to plasmas is that now the system of electrons under study is spatially strongly inhomogeneous -- due to the confinement by the potential $v$ of the atomic nucleus. To solve this problem the two-time correlation functions (\ref{ggtls_def}) were expanded into a basis of single-particle states or Hartree-Fock orbitals. This approach was successfully tested for ``artificial atoms'', i.e. electrons in quantum dots in refs.~\cite{balzer_prb09}. More recently it was applied by the present authors to electrons in small atoms \cite{hochstuhl_physe09} such as a one-dimensional beryllium model atom following earlier work by van Leeuwen and co-workers \cite{dahlen07}. In these works 
it was found that the ground state of few-electron atoms is well reproduced with simple approximations for the selfenergy such as the second Born approximation. The situation is more complex for time-dependent calculations. While lowest order calculations -- using a time-dependent Hartree-Fock selfenergy -- are well feasible \cite{hochstuhl_physe09} with the inclusion of correlations, simulations could only be carried out for very short times. This 
is a particular problem of ionization dynamics where one has to simultaneously treat bound and free electron states which requires large basis sets.

On the other hand, the interaction of atoms with a strong laser field can, in many cases, be assumed to be a coherent process. Further, at sufficiently low density, the interaction between atoms can be neglected. Also, at room temperature conditions the density matrix of the atom is dominated by the contribution of the ground state wave function, and one can instead of quantum statistics apply a pure state description.

\subsection{Wave function based methods}
At the conditions described above the initial state $|\psi_0\rangle$ of an atom is given by the ground state wave function or a coherent superposition of atomic eigenfunctions. Then the dynamics following an external exciation is described by the time-dependent $N$-electron Schr\"odinger equation (TDSE) with the hamiltonian (\ref{h})
\begin{equation}
i\frac{\partial}{\partial t}|\psi(t)\rangle = {\hat H}(t) |\psi(t)\rangle, \qquad \quad
|\psi(t=0)\rangle = |\psi_0 \rangle.
\end{equation}
The TDSE has been successfully solved in recent years by direct integration, exact diagonalization or various other techniques. For some recent results for atomic ionization with ultraviolet fields see \cite{Haan1994,bauch_pra08,krasovski_prl07} and references therein.
Yet these kinds of exact caculations are presently limited to $3\dots 6$ electrons, depending on the dimensionality. Therefore, approximate methods have been developed. A particularly successful approach is the {\em Multiconfiguration time-dependent Hartree-Fock} (MCTDHF) method due to Cederbaum \cite{Cederbaum_90}, see also \cite{Cederbaum_unified} and references therein. 
We have, therefore, implemented the MCTDHF method for photoionzation of atoms in a strong field and briefly outline the main idea. Starting point is the 
following ansatz for the many-electron wave function as linear combination of time-dependent Slater determinants which are denoted in terms of the 
occupation numbers $n_1, \dots n_M$ of the single-particle orbitals
\begin{equation}
 \ket{\psi(t)} \ = \ \sum_{\mathbf n} \; C_{\mathbf n}(t) \; \ket{n_1 , n_2 , \cdots , n_M ;t} \,.
\end{equation}
The sum runs over an integer number $\binom M N$ of determinants. In the limit $M \to \infty$ the method yields the exact wave function, whereas for 
$M=N$ one recovers the time-dependent Hartree-Fock approximation (TDHF). Thus, by increasing $M$ one can systematically improve the treatment of electronic correlations. The equations of motion for the coefficients $C_{\mathbf n}$ and the single-particle orbitals $| \phi_a \bigr \rangle $ are obtained by a variational principle and read, with a matrix notation (with respect to Slater determinants)
\begin{eqnarray}
\label{eq_wavefunction} 
i \, \dot{\mathbf C}(t) \ &=& \ \mathbf H(t) \, \mathbf C(t) \,,\\
\label{eq_orbitals} 
i \, \bigl | \dot \phi_a \bigr \rangle \ &=& \ \left( 1 -  \sum_r^M  \ket{\phi_r}\bra{\phi_r} \right) \; \left[ \; \hat h \, \bigl |  \phi_a \bigr \rangle \ + \ \sum_{ijkl}^M
\left\{ \boldsymbol \rho^{-1} \right\}_{ai} \rho_{ijlk} \, \hat W_{jk} \, \bigl |  \phi_l \bigr \rangle \; \right ]\,,	
\end{eqnarray}
%
where $\rho_{ij}$ and $\rho_{ijkl}$ denote the single-particle and two-particle density matrix, and $\hat h$ is the single-particle hamiltonian. Further,  $\hat W_{jk}$ is the operator of the mean-field potential which, in coordinate representation, is given by
 $W_{jk}(\mathbf r)= \int \; d \mathbf{\bar r}  \; \phi^\ast_j(\mathbf {\bar r}) \, \frac 1{|\mathbf r - \mathbf {\bar r}|} \, \phi_k(\mathbf {\bar r})$.
Eqs. (\ref{eq_wavefunction},\ref{eq_orbitals}) show that the expansion coefficients $\mathbf C$ obey a time-dependent Schr\"odinger equation, whereas the single-particle orbitals $|\phi_a\rangle$ are the solutions of a more complicated non-linear integro-differential equation.
\begin{SCfigure}\label{be_ion0}
 \includegraphics[width=28.5pc]{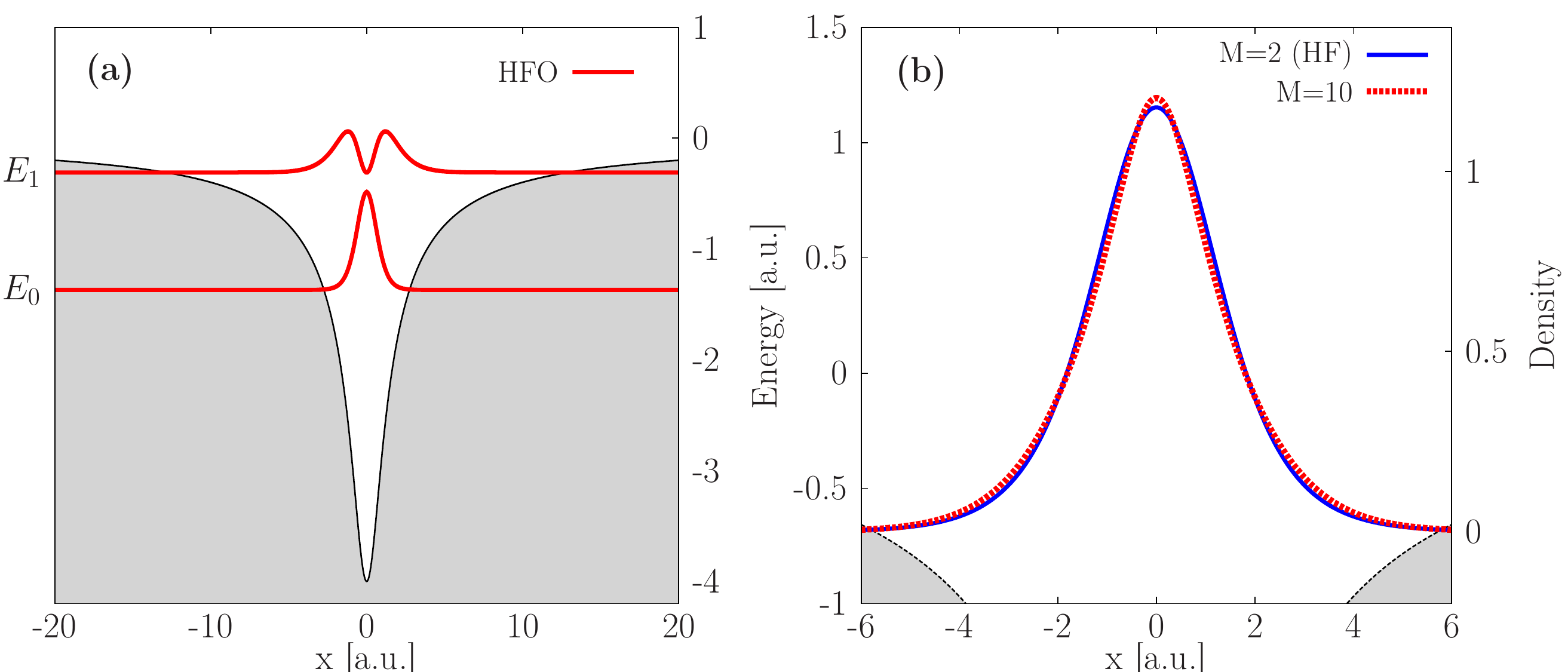}
 \caption{Sketch of the 1D Beryllium model atom. (a): Energies of the lowest Hartree-Fock orbitals and corresponding probability densities (vertically shifted). (b): total electron density prior to action of the laser pulse in Hartree-Fock (full blue line) and for $M=10$ (dashed red line with higher peak).}
\end{SCfigure}

\section{MCTDHF Results and Discussion}\label{s:results}
We now present numerical results for photoionization of a four-electron one-dimensional model atom (beryllium, $N=Z=4$) for which we presented nonequilibrium Greens functions results in ref.~\cite{hochstuhl_physe09}. For the $1D$ simulations we use regularized Coulomb potentials, $v(x) \ = \ - Z(x^2+\kappa^2)^{-1/2}$, and $w(x_1,x_2)\ = \ [(x_1-x_2)^2+\kappa^2]^{-1/2} \ $ where the screening parameters are introduced to avoid divergencies. This model has been used successfully in many studies of atom-laser interaction where also the influence of $\kappa$ has been investigated. Following refs.~\cite{Haan1994,hochstuhl_physe09} we use $\kappa=1$. 

A sketch of the atomic core potential and the (square of the) occupied HF basis functions (shifted by the orbital energies) is shown in Fig.~\ref{be_ion0} together with the total electron density. There are two bound states at the energies $E_0=-1.371$ and $E_1=-0.312$, each occupied by two electrons. The first task is to obtain accurate results for the ground state. The results of our MCTDHF calculations for the ground state energy are presented in the table. 
Convergence with respect to the correlation energy is reached for $M=9$. In comparison, the ground state energy for the same system from NEGF calculations in second Born approximation is found to be equal to $E_{2B} = -6.7713$, i.e. on the level of MCTDHF with $M=3$.
\begin{table}[h]
 \begin {center}
  \begin{tabular}{  c | c | c | c | c | c | c | c | c | c}
  \hline
  \hline
  \parbox[0pt][1.6em][c]{0cm}{} $\phantom{xxx} M \phantom{xxx}$ &  $\phantom{xxx} 2 \phantom{xxx}$ & $\phantom{xxx} 3 \phantom{xxx}$ & $\phantom{xxx} 4 \phantom{xxx}$ & $\phantom{xxx} 5 \phantom{xxx}$ & $\phantom{xxx} 6 \phantom{xxx}$
& $\phantom{xxx} 8 \phantom{xxx}$ 
& $\phantom{xxx} 10 \phantom{xxx}$\\
  \hline
  \parbox[0pt][1.6em][c]{0cm}{} Energy [Hartree]& $-6.7395$ & $-6.7713$ & $-6.7800$ & $-6.7829$ & $-6.7847$ 
& $-6.7850$ 
& $-6.7851$\\
  \hline
  \hline
  \end{tabular}
 \end{center}
\caption{Ground state energy of the 1D Beryllium atom with $\kappa=1$ for different numbers of determinants $M$. We used a basis with 50 eigenfunctions of the single-particle hamiltonian which assured convergence with respect to the basis size.}
\end{table}

\begin{SCfigure}\label{be_ion}
  \includegraphics[width=28.5pc]{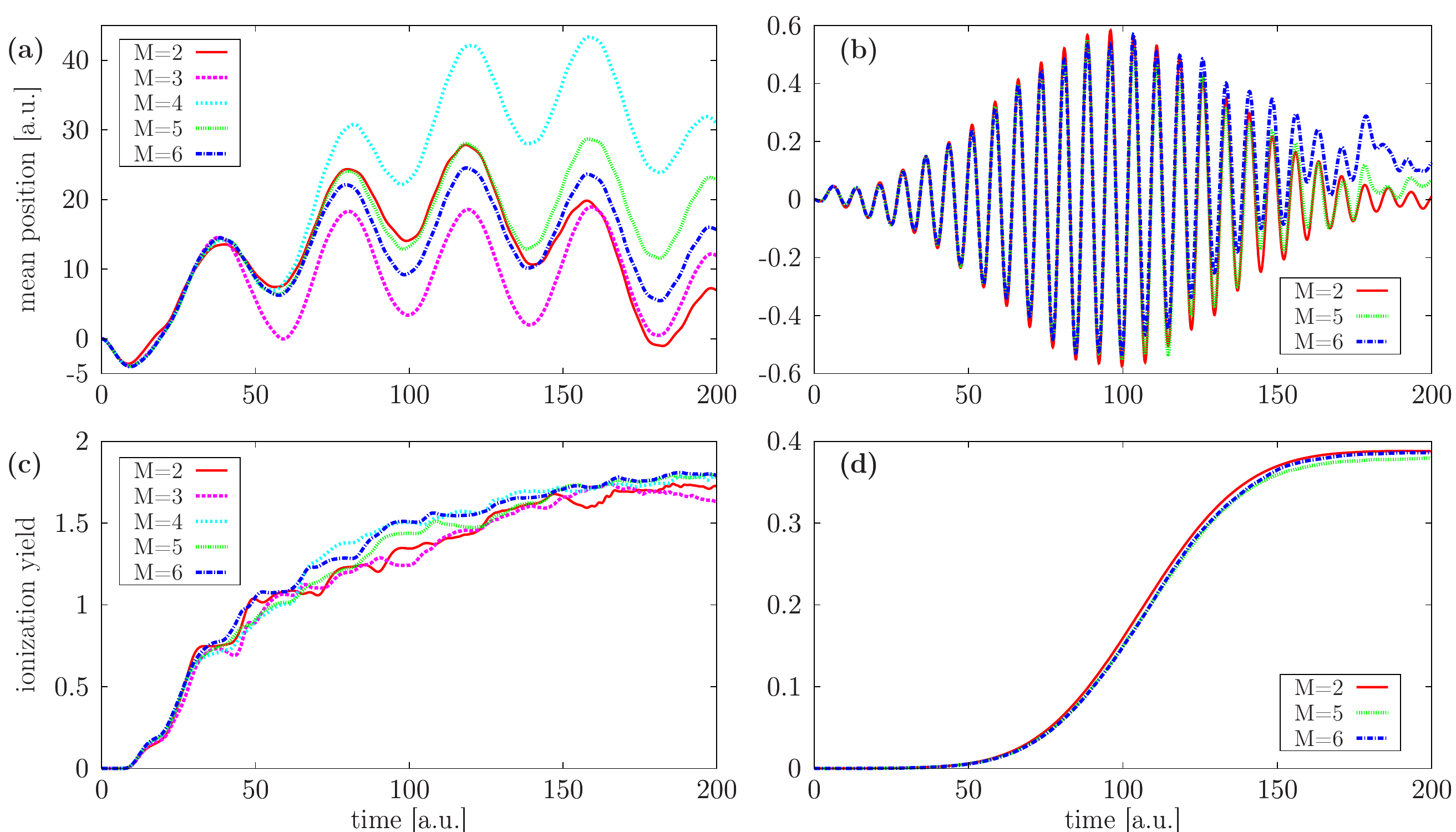}
  \caption{(Color online) Ionization dynamics of a 1D Beryllium model atom for two laser frequencies and different approximations for the electronic correlations.	Left column: $\omega=0.156 \,a.u$., Right column: $\omega=0.842 \, a.u$. Top row shows the mean electron coordinate and bottom row the mean number of ionized electrons.
}
\end{SCfigure}

As a next step, we use the ground state wave function as the initial state and expose the atom to a time-dependent laser field of the form (\ref{eq:gauss}) 
with a fixed number of cycles choosing the pulse width according to $\tau=10\pi/\omega$ and an amplitude of ${\cal E}_0=0.1$. We consider two frequencies, 
$\omega_1 = |E_1|/2$ and $\omega_2 = (|E_0|+|E_1|)/2$ which corresponds to a Keldysh-parameter $\gamma = \sqrt{I_p/2 U_p}$, [$I_p$ is the ionization potential, and $U_p$ the ponderomotive potential, $U_p = \mathcal{E}_0^2/4\omega^2$] of $\gamma_1=1.25$ and $\gamma_2=6.7$. In the first case we expect that direct photoionization is not possible and the dominant process will be tunnel ionization. In contrast, in the second case multiphoton ionization should be expected.
The time-dependent numerical results of our MCTDHF calculations are shown in Fig.~\ref{be_ion}. In the lower row we show the ionization yield [time-dependent expectation value of the number of ionized electrons $N_{ion}(t)$]. Interestingly, $N_{ion}(t)$ is much larger for the smaller photon energy (case 1) which confirms our previous results \cite{hochstuhl_physe09}. This is due to the much larger electron excursion amplitude in the low frequency laser field [recall that, for a free electron, the amplitude is $x_0\sim \omega^{-2}$]. For the larger photon frequency $\omega_2$, $x_0$ is strongly reduced. Nevertheless, the two electrons in the upper state with energy $E_1$ are driven substantially away from the nucleus during one half-cycle of the field but they are driven back during the following half-cycle. 
As a consequence, the ionization probability is substantially reduced. In this second case, MCTDHF calculations with different numbers of determinants $M$ are very close to each other, and even a TDHF calculation yields the qualitatively correct result. More interesting is the behavior for the smaller frequency. Here two electrons are driven far away from the nucleus and they occupy continuum states which leads to strong correlation effects. As a result, TDHF calculations -- while overall giving the correct picture -- become inaccurate already after about one laser cycle ($t \ge 40$).

Summarizing, we have presented MCTDHF calculations for a four electron model atom exposed to strong laser pulses with two different frequencies. These calculations are rather efficient and allow to follow the time-resolved ionization dynamics and the history of electronic correlations in single atoms. 
They show that, overall, TDHF gives the correct tendencies. However, it fails to accurately describe the dynamics of electrons at longer times, 
in particular in cases of strong ionization. 
In future work we will concentrate on correlation dynamics and on the inneratomic relaxation as well as on further comparisons with Nonequilibrium Greens functions results.

\section*{Acknowledgements}
This work is supported by the Deutsche Forschungsgemeinschaft via SFB-TR 24 and by the U.S. Department of Energy award DE-FG02-07ER54946.


\begin{thebibliography}{23}

\bibitem{Bauer2005} M.~Bauer, J. Phys. D {\bf 38}, R253 (2005)

\bibitem{Drescher2001} M.~Drescher et al.,
Science {\bf 291}, 1923 (2001)

\bibitem{Corkum2007} P.~B.~Corkum and F.~Krausz, 
Nature Phys. {\bf 3}, 381 (2007)

\bibitem{Uiberacker2007} M.~Uiberacker et al., Nature (London) {\bf 446}, 627 (2007)


\bibitem{krausz2009} F.~Krausz, and M.~Ivanov,
Rev. Mod. Phys {\bf 81}, 163 (2009)

\bibitem{kremp-etal.99pre} D. Kremp, Th. Bornath, M. Bonitz, and M. Schlanges, 
Phys. Rev. E {\bf 60}, 4725 (1999)

\bibitem{bonitz-etal.99cpp} M. Bonitz, Th. Bornath, D. Kremp, M. Schlanges, and W.D. Kraeft, 
Contrib. Plasma Phys. {\bf 39}, 329 (1999)

\bibitem{haberland-etal.01pre} H. Haberland, M. Bonitz, and D. Kremp, 
Phys. Rev. E {\bf 64}, 026405 (2001)

\bibitem{dahlen07} N.E.~Dahlen, and R. van Leeuwen, Phys. Rev. Lett. {\bf 98}, 153004 (2007)

\bibitem{hochstuhl_physe09} D.~Hochstuhl, K.~Balzer, S.~Bauch, and M.~Bonitz, 
Physica E (2009) in press, preprint arXiv: 0902.0768

\bibitem{bonitz-book} M. Bonitz, {\em Quantum Kinetic Theory}, Teubner, Stuttgart, Leipzig 1998.

\bibitem{kb-book} L.P. Kadanoff, and G. Baym, {\em Quantum Statistical Mechanics}, W.A. Benjamin, New York 1962.

\bibitem{kremp-book} D. Kremp, M. Schlanges, and W.-D. Kraeft, {\em Quantum Statistics of Nonideal Plasmas}, Springer, Berlin 2005.

\bibitem{bonitz-etal.jpcm96} M.~Bonitz et al., 
J. Phys.: Condensed Matter {\bf 8}, 6057 (1996)

\bibitem{bonitz97} M.~Bonitz, D.~Semkat, and D.~Kremp, Phys. Rev. E  {\bf 56},  1246 (1997);
D.~Semkat, D.~Kremp, and M.~Bonitz, Phys. Rev. E  {\bf 59},  1557 (1999)

\bibitem{kwong-etal.00prl} N.H. Kwong, and M. Bonitz, 
Phys. Rev. Lett. {\bf 84}, 1768 (2000)

\bibitem{lipavski} P. Lipavsky, V. Spicka, and B. Velicky, Phys. Rev. B {\bf 34}, 6933 (1986)

\bibitem{semkat-etal.05jp} D. Semkat, D. Kremp, and M. Bonitz, 
J. Phys.: Conf. Ser. {\bf 11}, 25 (2005)

\bibitem{kremp-etal.06jp} D. Kremp, D. Semkat, Th. Bornath, M. Bonitz, M. Schlanges, and P. Hilse, 
J. Phys.: Conf. Ser. {\bf 35} (2006), 53

\bibitem{balzer_prb09} K.~Balzer, M.~Bonitz, R. van Leeuwen, N.E.~Dahlen, and A.~Stan,
Phys. Rev. B {\bf 79}, 245306 (2009); K.~Balzer, and M.~Bonitz, 
J. Phys. A {\bf 42}, 214020 (2009),


\bibitem{Haan1994}
S.L. Haan, R. Grobe, and J.H. Eberly, Phys. Rev. A \textbf{50}, 378 (1994)

\bibitem{bauch_pra08} S.~Bauch, and M.~Bonitz, 
Phys. Rev. A {\bf 78}, 043403 (2008)


\bibitem{krasovski_prl07} E.E.~Krasovskii, and M.~Bonitz,
Phys. Rev. Lett. {\bf 99}, 247601 (2007) and submitted to Phys. Rev. A, preprint ArXiv: 0907.5384




\bibitem{Cederbaum_90} H.D.~Meyer, U.~Manthe, and L.S.~Cederbaum, J. Chem. Phys. {\bf 165}, 73 (1990)

\bibitem{Cederbaum_unified} O.E. Alon, A.I. Streltsov, and L.S. Cederbaum, 
J. Chem. Phys. {\bf 127}, 154103 (2007)


\end{thebibliography}
\end{document}